\documentclass[aps,prl,twocolumn,longbibliography,amsfonts,citeautoscript,floatfix,maxnames=4,10pt]{revtex4-2}
\usepackage[utf8]{inputenc}
\usepackage[english]{babel}
\usepackage{graphicx}
\usepackage{float}
\usepackage{bm}
\usepackage{xcolor}
\usepackage{epstopdf}
\usepackage{amsmath} 
\usepackage{amssymb}
\usepackage{epstopdf}
\usepackage{multirow}
\usepackage[normalem]{ulem} 
\usepackage{enumerate}
\usepackage{cases}
\usepackage{tikz}

\usepackage[urlcolor=blue,colorlinks=true,citecolor=red,linkcolor=red,pdfstartview={FitH},bookmarks=false]{hyperref}

\newcommand\doublecheck{\textcolor{black}{\checked\kern-0.6em\checked}}
\newcommand{\bs}{\boldsymbol}
\newcommand{\paren}[1]{\left( #1 \right)}
\newcommand{\sqbrack}[1]{\left[ #1 \right]}

\graphicspath{{figs/}{./figs/}{.}}
\usepackage{lipsum}

\newcommand{\DeltaSO}{{\Delta_{\rm so}}}

\newcommand{\Tr}{\text{Tr}}

\begin{document}

\newcommand{\JYU}{\affiliation{Department of Physics and Nanoscience Center, University of Jyväskylä, P.O. Box 35 (YFL), FI-40014 University of Jyväskylä, Finland}}

\title{
Detection of spin- and valley-polarized states in van der Waals materials via thermoelectric and non-reciprocal transport}

\author{Oladunjoye A. Awoga}
\email[e-mail:]{oladunjoyea.a.awoga@jyu.fi}

\author{Pauli Virtanen}

\author{Tero T. Heikkil\"a}

\author{Stefan Ili\'c}

\JYU

\begin{abstract}
We predict thermoelectric and current rectification effects in hybrid junctions formed by Ising superconductors and materials hosting valley-polarized states. Both effects originate from the interplay of intrinsic Ising spin-orbit coupling, spin-splitting from an exchange or Zeeman field, and valley polarization. The resulting transport signatures provide experimentally accessible probes of valley-polarized states in van der Waals heterostructures, such as junctions of few-layer transition metal dichalcogenides and twisted bilayer or rhombohedral graphene. 
\end{abstract}

\maketitle
\emph{Introduction.-}\label{sec.int} Since the discovery of graphene, two-dimensional (2D) materials have become a central platform for exploring emergent quantum phenomena. The ability to assemble atomically thin layers into van der Waals heterostructures has enabled unprecedented control over system properties and engineering of novel quantum states \cite{novoselov20162d}. A common feature in many of these materials is the presence of multiple inequivalent band-structure extrema in the momentum space, known as \emph{valleys}, an essential electronic degree of freedom for their low-energy description. Exploiting valleys for electronic functionality defines the field of valleytronics \cite{Rycerz.2007.Valley, schaibley2016valleytronics}. Therefore, understanding how valley structure shapes correlated and superconducting states has become a major theme in van der Waals materials.

 A prime example of valley-dependent physics is found in few-layer transition metal dichalcogenides (TMDs) \cite{zhu2011giant,xiao2012coupled, kormanyos2015k}. The absence of inversion symmetry in these materials enables a strong intrinsic spin-orbit coupling (SOC), which pins the electron spins to the out-of-plane direction, with opposite orientations in the two valleys. This spin-valley locking has remarkable consequences for intrinsic superconductivity in TMDs. Cooper pairs in TMDs have an unusual structure: they are formed by electrons from opposite valleys whose spins are strongly locked out-of-plane. This is known as Ising superconductivity, whose main hallmark is great robustness to in-plane magnetic field, which was experimentally confirmed in various superconductig TMDs \cite{lu2015evidence, saito2016superconductivity, xi2016ising, xing2017ising, lu2018full, de2018tuning}. 

More recently, valley physics plays a central role in the correlated states of graphene-based systems~\cite{cao2018unconventional, cao2018correlated, po2018origin, peltonen2018mean, yankowitz2019tuning, sharpe2019emergent, serlin2020intrinsic, nuckolls2020strongly, bultinck2020mechanism, xie2020nature, zhou2021superconductivity, zhou2021half, chatterjee2022inter,Pangburn2023Superconductivity, arp2024intervalley}. In twisted bilayer graphene, a small relative twist between graphene layers produces a long-wavelength moire superlattice that strongly influences the band structure, leading to flat bands and enhanced interaction effects~\cite{bistritzer2011moire}. These conditions stabilize superconductivity and correlated insulating phases, as well as valley-polarized states in which electrons preferentially occupy one of the two valleys \cite{ cao2018unconventional, cao2018correlated, serlin2020intrinsic, nuckolls2020strongly}. In rhombohedral (ABC-stacked) graphene, the enhanced low-energy density of states at the surface layers \cite{zhang2010band,Kopnin2013High,Awoga2024Superconductivity} amplifies interactions and can instead drive the system into a quarter-metal phase (QM) \cite{zhou2021half}, where the low-energy electronic states are polarized in both spin and valley.  

So far, valley physics has been experimentally explored primarily using optical probes~\cite{mak2012control, zeng2012valley, cao2012valley}, high-field magnetotransport measurements revealing quantum Hall states~\cite{feldman2009broken, du2009fractional, young2012spin}, and nonlocal transport measurements where  Berry-curvature-related effects  are believed to play a role~\cite{gorbachev2014detecting, sui2015gate, mak2014valley}. While these methods have provided important insights, straightforward electrical transport probes of valley-dependent phenomena, particularly those that are compatible with superconducting or hybrid van der Waals structures,  are still lacking. 

 %
\begin{figure}[!t]
    \centering
        \includegraphics[width=0.45\textwidth]{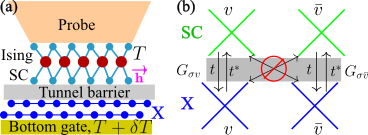}
	 \caption{Schematic illustration for probing Ising thermopower and rectification. (a) Junction between Ising SC and a spin- and valley-polarized material X with a tunnel barrier. The field $\boldsymbol{h}$, in the SC can be applied with an in-plane external field or via proximity effect with a ferromagnetic insulator. The probe can be an STM tip on which the SC is grown or can be a gate for tuning the superconducting phase, while the bottom gate tunes the electronic state of X. The thermoelectric signal results from a temperature difference, $\delta T$, between the materials, while rectification occurs as a consequence of voltage differences beyond the linear regime.   
      (b) The valley and spin dependence of the tunneling. Only intra-valley tunneling with tunneling amplitude $t$ of the same spin species is allowed. However, due to the valley and spin polarization in X, the tunneling conductance $G_{\sigma v}^{}$ inherits spin, $\sigma$, and valley, $v$, selectivity enabling the probing of valley- and spin-polarized states in the Ising SC.
    }
	\label{fig:Schematic}
\end{figure}
In this Letter, we predict thermoelectric and current rectification effects that provide such direct transport probes in hybrid junctions combining an Ising superconductor (ISC) with a spin-valley-polarized material. Applying an in-plane Zeeman field generates valley-odd triplet correlations in the ISC, which then couple to the spin-valley polarized states in the adjacent material. This process breaks electron-hole symmetry and thereby produces a thermoelectric response \textendash {\em Ising thermopower}\textendash, alongside a current rectification effect. These complementary transport signatures can be used to experimentally detect spin-valley polarized states.
%
%

\emph{The setup and model.-} 
We consider the setup shown in Fig.~\ref{fig:Schematic}(a), where an ISC is tunnel-coupled through a tunnel barrier to a spin- and valley- polarized material (for example a QM). The barrier is assumed to be spin- and valley-independent. More generally, the same effects arise in other junctions as long as both spin and valley polarizations are present in the system, either in X or in the barrier, as discussed in End Matter.

The model assumes well-defined valley-conserving
tunneling across the tunneling interface. It can be realized using a flat extended contact, for example in a scanning tunneling microscopy (STM) geometry with a planar tip that forms a 2D tunnel junction, as shown in  Fig.~\ref{fig:Schematic}(a). Then, momentum and valley-conserving tunneling occurs when the Fermi surfaces of the two materials overlap, as sketched in Fig.~\ref{fig:Schematic}(b). Our model requires only valley conservation, which is a weaker constraint, since intra-valley scattering does not modify our results. This type of setup has recently been used in the quantum twisting microscope~\cite{Inbar2023the}, providing a powerful approach to probing various properties of van der Waals materials~\cite{Birkbeck2025quantum}, including TMDs~\cite{Pichler2024Probing}.

\begin{figure}[!t]
    \centering
       \includegraphics[width=0.30\textwidth]{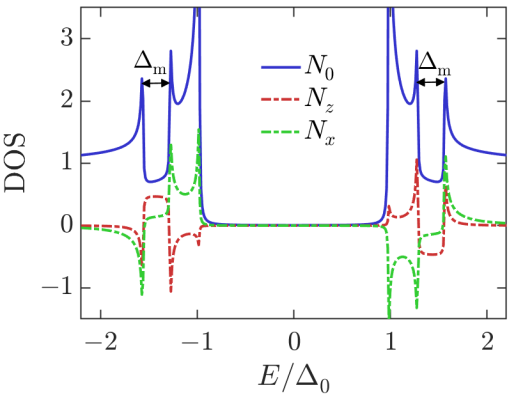}
	\caption{Various DOS in Ising SC with in-plane field showing both energy and spectral gap, $\Delta_{\rm}^{}$, symmetries. These DOS are responsible for different transport coefficient, see Eq.~\eqref{eq:Transport-coeffs}. Here, $ \Delta_{\rm so}^{}/\Delta_0^{}=1,\, h_{\rm }^{}/\Delta_0^{}=0.2,\, \psi = 0$, and $\Delta_0^{} \equiv \Delta \paren{T=0,h=0}$.}
	\label{fig:DOS}
\end{figure}

%
%
The ISC is described by the Bogoliubov-de-Gennes Hamiltonian~\cite{Ilic.2023.Spectral} ,
\begin{equation}\label{eq:Ising_BdG}
    \hat H\left(\bs q\right)= \left(\xi_{\bs q}^{} +  \Delta_{\rm so}\,  \hat s_z^{} \hat \eta_z^{}\right)\, \hat \tau_z^{} +  h\, \hat s_x^{} - \Delta  \hat\tau_x^{} +\psi \hat s_y  \hat\tau_y^{}.
\end{equation}
 Here, $ \hat s_i^{},\, \hat\tau_i^{}$ and $ \hat\eta_i^{}$ are, respectively, the $i$th-Pauli matrices in spin, Nambu and valley spaces, $\xi_{\bs q}^{},\,\DeltaSO$, and $ h$ represent, respectively, the normal-state single-particle excitation energy, the Ising spin-orbit strength, and the in-plane exchange field. Superconductivity is captured by the order parameters $\Delta$, corresponding to the conventional $s$-wave singlet component, and $\psi$, an $f$-wave triplet component predicted theoretically \cite{mockli2019magnetic, mockli2019magnetic}.  Next, it is useful to define the spectral function of the ISC, $\hat{A}^S\paren{E} =-\frac{1}{\pi}\Im\sqbrack{\sum_{\bs q} \paren{E+i\, 0^+ - \hat{H}\paren{\bs q}}^{-1}}$, where $E$ is the energy. The structure of this spectral function is
 \begin{equation}\label{eq:Ising-GF}
\begin{split}
  \hat A^{\rm S}\paren{E}  =  N_{0}^{\rm } + N_{x}^{\rm }  \hat s_x^{}   +  N_{z}^{\rm }  \hat s_z^{}\, \hat \tau_z^{}\, \hat \eta_z^{} + \hat{F} 
  \end{split}
\end{equation}
where the spectral terms $N_{j}^{}$ are explicitly given in Eq.~\eqref{eq:App_Ising_GF} in the End Matter and shown in Fig.~\ref{fig:DOS}. For single particle tunneling, only these spectral terms are relevant, so the anomalous (pairing) terms $\hat{F}$ are omitted. The conventional spin-averaged density of states (DOS) is given by $N_0^{}$. It exhibits the characteristic features of an ISC in an in-plane field known as mirage gaps, denoted by $\Delta_{\rm m}^{}$~\cite{tang2021magnetic,Ilic.2023.Spectral}. The second contribution, $N_x^{}$, corresponds to the spin-odd component of the DOS along the applied field direction. Such a term arises generically in superconductors subjected to Zeeman fields, reflecting the energy shifts between opposite spin species. The presence of SOC diminishes this component and ultimately suppresses it at very strong SOC. Finally and most importantly,  ISCs host an additional component unique to their coupled spin and valley structure, $N_{z}^{}$, which we refer as Ising DOS. This contribution is odd under exchange of valleys and originates from the interplay between Ising SOC and the applied field. As we show below, this valley-odd component is responsible for the thermoelectric effect and rectification reported in this Letter.  As shown in Fig.~\ref{fig:DOS}, $N_0^{}$ is even in energy while both $N_z$ and $N_x$ are odd in energy. 
%
%

%
%
The minimal model for X, that is, a material hosting both spin and valley polarized states, is given by the Hamiltonian
\begin{equation}\label{eq:Hamiltonian-QM}
    H_{\rm X}^{} = \xi_{\bs p}^{}\, \hat \tau_z^{} +   m_{\rm v}^{}\,\hat  \eta_z^{} + m_{\rm z}^{}\,\hat  s_z^{} +   m_{\rm x}^{}\,\hat  s_x^{}
\end{equation}
where $m_{\rm z}^{}$ and $m_{\rm x}^{}$ denote, respectively, out-of-plane and in-plane spin magnetization,  and $m_{\rm v}^{}$ is valley polarization. Such spin and valley polarized states can arise in rhombohedral graphene multilayers, where interaction-driven symmetry breaking lifts both spin and valley degeneracies~\cite{zhou2021half}. The direction of spin magnetization may depend on additional details of these systems, such as SOC and geometry, so for generality we consider both $m_{\rm z}^{}$ and $m_{\rm x}^{}$, taking one of them finite at a time.   

Finally, the tunnel barrier between the ISC and X is modeled by the tunneling Hamiltonian
$H_{\rm T}^{} = t\, \hat \tau_z^{}$, 
where $t \in \mathbb{R}$ is the tunneling amplitude, shown schematically in Fig.~\ref{fig:Schematic}(b).

\emph{Charge and heat currents.} We proceed to calculate the charge ($I \equiv I_{q=0}^{}$) and heat ($\dot Q_{}\equiv I_{q=1}^{}$)  currents, following the standard tunneling approach~\cite{mahan.2000.many-particle,Ozaeta.2014.predicted,IsingSM},
\begin{equation}\label{eq:averaged-currents}
 I_{q}^{} =  \frac{\mathcal{G}_{\rm T}^{\rm }}{e}
 \int_{-\infty}^\infty {\rm d} E\, \paren{\mu -E}^q \sqbrack{N_{0}^{}  + \mathcal M_{\rm s}^{x}N_x^{} +  \mathcal  M_{\rm sv}^{z} N_{z}^{}} f_{\rm XS}^{}.
\end{equation}
We also calculate the spin, valley, and spin-valley currents, and their respective heat counterparts, presented for completeness in the End Matter. 
 We denote the normal-state tunneling conductance with $\mathcal{G}_{\rm T}^{}$, $\mu$ is the chemical potential, $f_{\rm XS}^{}=f_{\rm X}^{}-f_{\rm S}^{}$, and $f_i$ ($i={\rm X,S}$) the Fermi-Dirac distribution function with temperature $T_i$ and potential $\mu_i$. We introduced the spin polarization along the $x$ direction $\mathcal  M_{\rm s}^x$, and the spin-valley polarization along the spin $z$ direction $\mathcal  M_{\rm sv}^z$. Both polarizations are in the interval $[-1,1]$, where $\pm1$ describes perfect polarization, while $0$ indicates the absence of polarization. They stem from the non-zero spin and valley magnetizations in the X described by  Eq.~\eqref{eq:Hamiltonian-QM}, and can be estimated as $ \mathcal  M_{\rm s}^x\sim m_{\rm x}^{}/\mu$ and $M_{\rm sv}^z\sim m_{\rm z} m_{\rm v}/\mu^2$~\cite{IsingSM}. 
The current \eqref{eq:averaged-currents} has three distinct contributions, each associated with a different component of the DOS shown in Fig.~\ref{fig:DOS}. The term containing $N_0$, which is even in energy, yields the conventional (non-thermoelectric, non-rectifying) contribution to the current. On the other hand, terms containing $N_x$ and $N_z$ are both odd in energy and thus break electron-hole symmetry, so they couple to different-symmetry component of the distribution function difference $f_{\rm XS}^{}$, giving rise to thermoelectric and rectification effects.

\emph{Thermoelectric effects.} To make the thermoelectric effect more transparent, we next consider the linear-response regime, assuming that both the voltage bias $ V=(\mu_{\rm X}^{}-\mu_{\rm S}^{})/e$ and the temperature gradient $\delta T=T_{\rm X}^{}-T_{\rm S}^{}$ across the junction are small, and find the currents up to the first order in $V$ and $\delta T$. We obtain 
\begin{equation}\label{eq:Onsager}
   \begin{pmatrix}
    I   
        \\
        \dot{Q}
    \end{pmatrix}   = 
     \begin{pmatrix}
    G   & \mathcal  M_{\rm s}^{x}\,\alpha_{x}^{} + \mathcal  M_{\rm sv}^{z}\,\alpha_{z}^{}  
        \\
   \mathcal   M_{\rm s}^{x}\,\alpha_{x}^{} + \mathcal M_{\rm sv}^{z}\,\alpha_{z}^{}  & G_{\rm th} T
    \end{pmatrix}
    \,
    \begin{pmatrix}
         V
        \\
        \delta T/T
    \end{pmatrix}. \\ 
\end{equation}
 We define the following transport coefficients,
\begin{subequations}\label{eq:Transport-coeffs}
    \begin{align}
       G_{q }^{} & =-\frac{G_{\rm T}^{}}{e^{q}} \int_{-\infty}^{\infty} {\rm d}E\,\frac{E^q}{T^{\frac{q}{2}}}\,N_{0}^{}\paren{E}\, f_{\rm }^{\prime}\paren{E}, \ q=0,2 \label{eq:conductances}, 
       \\
         \alpha_{x/z}^{}  & = -\frac{G_{\rm T}^{}}{2e} \int_{-\infty}^{\infty} {\rm d}E\, E\, \,N_{x/z}^{}\paren{E}\, f_{\rm }^{\prime}\paren{E}\, \label{eq:TEs},
    \end{align}
\end{subequations}
where Eq.~\eqref{eq:conductances} describes the conventional  electrical conductance, $G_0^{} \equiv G$, and thermal conductance, $G_2^{}\equiv G_{\rm th}^{}$, whereas Eq.~\eqref{eq:TEs} describes the thermoelectric coefficients.
%
\begin{figure}[!t]
    \centering
        \includegraphics[width=0.47\textwidth]{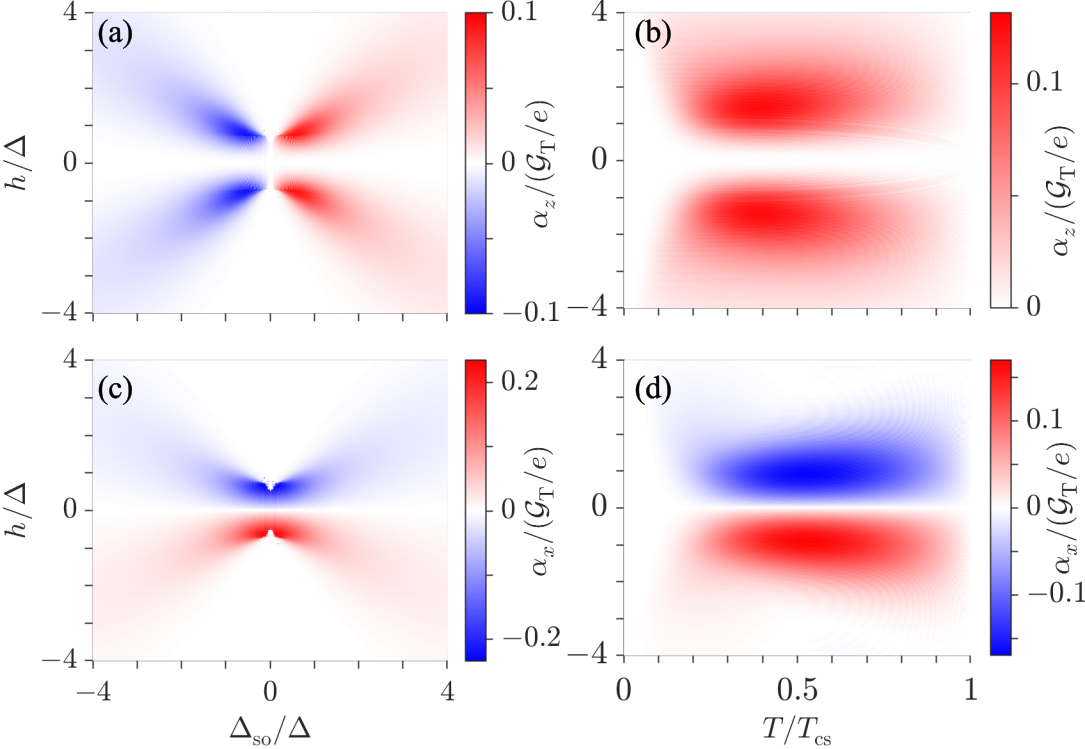}
	\caption{Thermoelectric coefficients in Ising SC. $\alpha_z^{}$ in $h-\Delta_{\rm so}$ space (a) and $h- T$ space (b). (c,d) Same as (a,b) but for $\alpha_x^{}$. Here $\psi=0$,\, $k_{B}^{}T/\Delta_0^{}=0.3$ in (a,c), and $\Delta_{\rm so}/\Delta_0^{} =1$ in (b,d).
    }
	\label{fig:TE-Coeffs}
\end{figure}

The thermoelectric coefficient $\alpha_z$ captures the Ising thermopower. It is one of the main results of our work, showing that the valley-odd correlations encoded in the Ising DOS $N_z$, which are unique to ISCs,  couple to spin-valley-polarized states in X, thus generating a thermoelectric response. The coefficient $\alpha_x$, on the other hand, represents another thermoelectric effect previously established \cite{Machon.2013.Nonlocal,Ozaeta.2014.predicted,Kolenda.Observation.2016,Bathen.2017.spin,Bergeret.2018.Nonequilibrium,Sun.2024.Josephson} in the context of spin-split superconductors. It arises from the combination of spin-split superconductor with a material that is spin-polarized in the same direction ($x$-direction in our case) as follows from Eq.~\eqref{eq:Transport-coeffs}.

In Fig.~\ref{fig:TE-Coeffs}, we plot both thermoelectric coefficients as a function of SOC strength, applied Zeeman field and temperature. The two coefficients reach comparable maximal magnitudes. A key distinction lies in their symmetry with respect to the applied field: $\alpha_x(-h)=-\alpha_x(h)$, whereas $\alpha_z(-h)=\alpha_z(h)$. This provides a way to experimentally distinguish the $\alpha_z$ effect, which signals the presence of spin-valley polarized states, from the more conventional $\alpha_x$ response. In the limit of large SOC and low temperatures, 
$\Delta_{\rm so}\gg \Delta\gg k_{\rm B}^{}T$,
we find analytically (see the End Matter)
\begin{equation}\label{Eq:ThermoAN}
\alpha_z\approx \frac{G_T}{e} \frac{h^2\tilde{\Delta}^2}{\Delta_{\rm so} (\Delta_{\rm so}^2+h^2)}\sqrt{\frac{\pi\tilde{\Delta}}{2k_{\rm B}^{} T}} e^{-\tilde{\Delta}/k_{\rm B}^{} T},
\end{equation}
and $\alpha_x=-\Delta_{\rm so}\alpha_z/h$, where we introduced the renormalized gap $\tilde{\Delta}=\Delta \Delta_{\rm so}/\sqrt{\Delta_{\rm so}^2+h^2}$. We see that large SOC suppresses the effects in a power-law manner, however, an appreciable effect can still be achieved in this regime with a sufficiently large exchange field $h$. Both thermoelectric effects require a finite temperature, since thermally activated quasiparticles are needed to enable transport at low biases. Finally, we note that the f-wave triplet component $\psi$ leads only to minor quantitative changes of these results, and we thus set $\psi=0$ throughout the Letter. For completeness, in the End Matter, we consider $\psi \neq 0$. 

\begin{figure}[!t]
    \centering
        \includegraphics[width=0.49\textwidth]{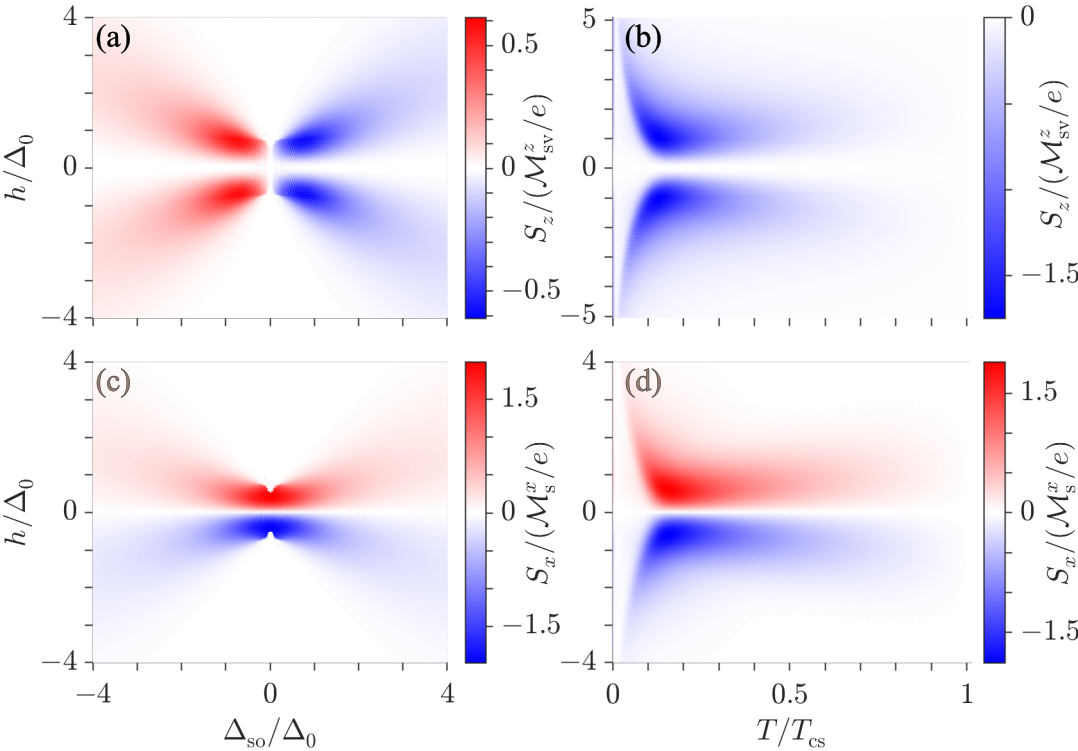}
	\caption{Seebeck coefficients from Ising SC, $S_{x/z}^{}$ corresponding to the TEs in Fig.~\ref{fig:TE-Coeffs}.  
    }
	\label{fig:Seebeck}
\end{figure}

While the thermally driven current is a direct manifestation of the thermoelectric effect, the quantity that is most commonly measured in experiments is the thermopower, or the Seebeck coefficient. It is defined as the voltage $V$ that develops in response to a temperature bias $\delta T$ when the circuit is open, so that the net current vanishes $(I=0)$. Using the transport coefficients from Eq.~\eqref{eq:Transport-coeffs}, we can define $S_z^{}=-\mathcal M_{\rm sv}^z\,\alpha_z^{}/GT$ and $S_x^{}=- \mathcal M_{\rm s}^x\,\alpha_x^{}/GT$. The corresponding plots are shown in Fig.~\ref{fig:Seebeck}, showing a qualitatively similar behavior to the thermoelectric coefficient. 

\emph{Rectification effects. }We now turn to non-linear transport, where additional consequences of electron-hole symmetry breaking become apparent, beyond thermoelectric effects. In particular, current rectification arises when the current has a voltage-symmetric component, leading to a non-reciprocal current-voltage relation $I(V)\neq -I(-V)$. Then, one bias direction is preferred over the other, corresponding to lower resistance. In other words, the system acts as a diode and rectifies AC signals.

To make this more explicit, let us decompose the total current \eqref{eq:averaged-currents} into three contributions, 
$I=I_0+\mathcal{M}_{\rm s}^x I_x+M_{\rm sv}^z I_z$, 
where
$I_i=\frac{\mathcal{G}_{\rm T}^{}}{e} \int {\rm d}E \, N_i f_{\rm XS}$.
The component $I_0$ is anti-symmetric in voltage, $I_0(-V)=-I_0(V)$, whereas the other two components are symmetric, $I_{x,z}(-V)=I_{x,z}(V)$, reflecting electron-hole symmetry breaking. The latter components therefore lead to current rectification. Similarly to the discussion of thermoelectric effects, we can identify two distinct sources or rectification. The first arises from $I_x$, and has been previously studied in Zeeman-split superconductors \cite{strambini2022superconducting, ilic2022current}. The second is the novel contribution from $I_z$, originating from Ising DOS and spin-valley polarization.

\begin{figure}[!t]
    \centering
        \includegraphics[width=0.49\textwidth]{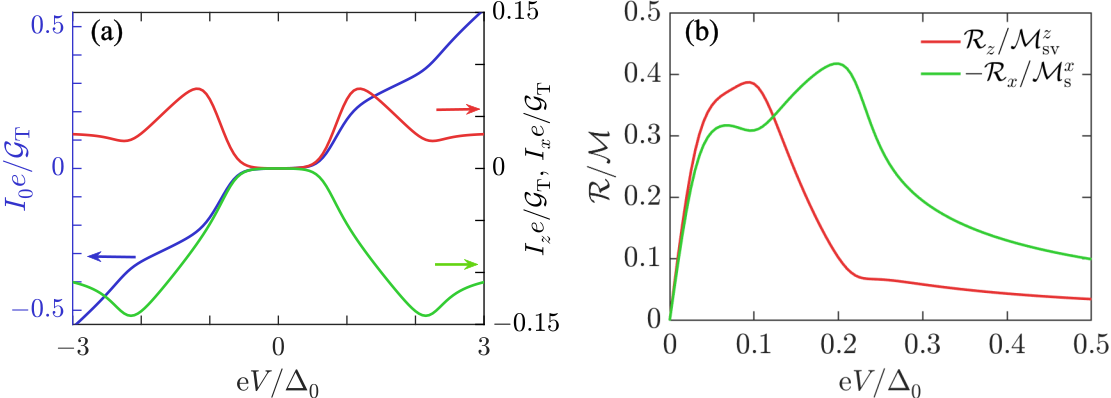}
	\caption{Rectification effects from Ising SC. (a) Individual currents, $I_{0,x,z}^{}$, and (b) rectification factors. Here $\Delta_{\rm so}/\Delta_0=h/\Delta_0=1$ and $k_{\rm B}^{}T/\Delta_0 = 0.1$.}
	\label{fig:Rect}
\end{figure}

We quantify the rectification effect by the coefficient 
\begin{equation} \label{Eq:Rect}
    \mathcal{R} =[I(V)+I(-V)]/[I(V)-I(-V)],
\end{equation}
which represents the ratio of voltage-symmetric to voltage-asymmetric component of the current. A larger $|\mathcal{R}|$ therefore corresponds to better rectification. Assuming that only one polarization, $\mathcal{M}_{\rm s}^x$ or $\mathcal{M}_{\rm sv}^z$, is finite at a time, we can define the coefficients describing the two rectification mechanisms separately
\begin{equation}
    \mathcal{R}_x =\mathcal{M}_{\rm s}^x I_x/I_0, \qquad \mathcal{R}_z =\mathcal{M}_{\rm sv}^z I_z/I_0.
\end{equation}
In Fig.~\ref{fig:Rect} we plot them, alongside the corresponding currents.

The two mechanisms behave differently under magnetic field reversal, namely $\mathcal{R}_x(-h)=-\mathcal{R}_x(h)$ and  $\mathcal{R}_z(-h)=\mathcal{R}_z(h)$, similarly to thermoelectric coefficients discussed above. Moreover, they show distinct voltage dependences, as shown in Fig.~\ref{fig:Rect}. These differences provide a clear way to distinguish the two mechanisms experimentally. In particular, measuring $\mathcal{R}_z$ directly probes the spin-valley polarized states, whereas $\mathcal{R}_x$ comes from more conventional spin-splitting. The magnitude of both effects is governed by corresponding polarizations, so that $\mathcal{R}_x\le\mathcal{M}_{\rm s}^x$ and $\mathcal{R}_z<\mathcal{M}_{\rm sv}^z$. The coefficient $\mathcal{R}_x$ is maximized at vanishing Ising SOC, where it can reach $\mathcal{R}_x=\mathcal{M}_{\rm s}^x$  at optimal voltages \cite{ilic2022current}, whereas $\mathcal{R}_z$ requires finite SOC and is maximized around $\Delta_{so}\sim \Delta$. At large SOC and low voltages $\Delta_{\rm so}\gg\Delta\gg eV$, we find analytically (see the End Matter)
\begin{equation}
\mathcal{R}_z (V) \approx -\mathcal{M}_{\rm sv}^z \frac{h^2\tilde{\Delta}}{\Delta_{\rm so}(\Delta_{\rm so}^2+h^2)} \tanh\paren{\frac{eV}{2k_{\rm B}^{}T}}
\end{equation}
$\mathcal{R}_x=-\Delta_{\rm so} \mathcal{M}_{\rm s}^x \mathcal{R}_z/(\mathcal{M}_{\rm sv}^z h)$.
These expressions have the same field and SOC dependence as the thermoelectric effects [see Eq.~\eqref{Eq:ThermoAN}]. From here, we see that the maximal rectification is achieved at voltages $eV\sim 4  k_{\rm B}^{} T$. 

The rectification effect provides a complementary probe of spin-valley polarized states along thermoelectric measurements, and both can be accessed within the same experimental setup. Moreover, rectification has an advantage of being a purely electrical probe, not requiring the temperature gradient across the structure. It can also be seen in AC measurements, where it generates DC response and second-harmonic signals.
%

\emph{Experimental detection.-} An optimal platform to detect the effects proposed here is a junction between an Ising superconductor, typically realized in few-layer TMDs, and a spin-valley polarized material such as rhombohedral graphene, forming an extended and clean interface as sketched in Fig.~\ref{fig:Schematic}(a). Alternatively, one may employ a system where material X only exhibits valley polarization, for instance twisted bilayer graphene at suitable gating. In that case, the tunneling barrier must be spin-active to ensure the missing ingredient of spin polarization. That could be achieved by using a ferromagnetic insulating layer polarized along the $z$-direction as the tunneling barrier (see End Matter). 

As shown in Figs.~\ref{fig:TE-Coeffs} and \ref{fig:Seebeck} and discussed in Fig.~\ref{fig:Rect}, a very strong Ising SOC is detrimental for both thermoelectric and rectification effects. This is a relevant consideration when selecting the TMD material, since the Ising SOC varies widely, from $\Delta_{\rm so}\sim$ a few $\Delta_0$ in gated MoS$_2$ monolayers \cite{saito2016superconductivity}, to $\Delta_{\rm so}\gg\Delta_0$ in WS$_2$ and NbSe$_2$ monolayers \cite{xi2016ising, lu2018full, de2018tuning}. Moreover, in multilayer samples the effective SOC is reduced due to partial cancellation in adjacent layers, vanishing in the bulk, but remaining finite in few-layer systems \cite{de2018tuning}. For maximal signal, therefore, it is preferable to use ${\rm MoS_2}$ or a multilayer TMD. Note that using MoS$_2$ imposes additional constraints, as superconductivity typically requires ionic liquid gating incompatible with STM, and therefore either alternative gating \cite{cao2023full,ji2024continuous} or a non-STM setup \cite{costanzo2018tunnelling} is required.

Another experimental concern is the presence of disorder. While Ising superconductivity is robust against intra-valley scattering, it is sensitive to inter-valley scattering, which mixes the two valleys and produces pair-breaking effects \cite{ilic2017enhancement}. The latter requires a large momentum transfer, and therefore it arises primarily from short-range defects such as vacancies or sample edges. Previous studies have shown that the essential spectral features, such as mirage gaps, persist as long as $\tau_{\rm iv}^{-1}<\Delta_0$ \cite{Ilic.2023.Spectral} . Thus, the predicted effects should remain observable in samples with minimal short-range defects. 

\emph{Summary.} In summary, we have shown that a junction of an Ising superconductor and a material hosting spin-valley polarized states gives rise to new types of thermoelectric and rectification effects. Their experimental observation would provide a direct transport probe of valley-dependent physics, which is relevant for valleytronics and for the rapidly growing field of interaction-driven valley-ordered and symmetry-broken states. We identify a junction of a few-layer TMD combined with a graphene-based system, such as twisted bilayer or rhombohedral graphene, as promising platforms to realize this proposal.

\begin{acknowledgments}
This work is part of the Finnish Centre of Excellence in Quantum Materials (QMAT). We acknowledge support from the Research Council of Finland (Projects No. 354735 and 355056) and through the Finnish Quantum Flagship, project number 359240. 
P.V. acknowledges support by European Union’s HORIZON-RIA programme 331 (Grant Agreement No.~101135240 JOGATE), and by Jane and Aatos Erkko Foundation, Keele Foundation, and Magnus Ehrnrooth Foundation as part of the SuperC collaboration.
\end{acknowledgments}
%
\bibliography{biblioQuasiclassical}
\clearpage
\appendix
\renewcommand{\theequation}{A\arabic{equation}}
\setcounter{equation}{0}

\renewcommand{\thefigure}{A\arabic{figure}}
\setcounter{figure}{0}
\section*{END MATTER}

\subsection*{Derivation of currents}
The setup contains mainly three parts, namely Ising superconductor (ISC), barrier, and material X with corresponding Hamiltonian $H$, $H_{\rm T}^{}$  and $ H_{\rm X}^{}$, respectively, all described in the main text. Thus, the total Hamiltonian is $H + H_{\rm T}^{} + H_{\rm X}^{}$. Here, we derive the tunneling current for a general X/barrier/ISC junction. We then present results for the ISC/tunnel barrier/Spin-valley-polarized X junction (discussed in the main text) and ISC/spin-filtering barrier/Valley-polarized X junction. To fully understand the tunneling current, we first discuss each section of the setup.
%
%
%
\subsection{Ising superconductor with an in-plane magnetic field}\label{App:sec:IsingSC}
The Hamiltonian of an ISC with an applied in-plane field is given in Eq.~\eqref{eq:Ising_BdG} and the corresponding spectral function is given in Eq.~\eqref{eq:Ising-GF} in the main text. The spectral functions are~\cite{Awoga.2026.Nonequilibrium},
\begin{subequations}\label{eq:App_Ising_GF}
\begin{align}
   N_{0}^{\rm } &=  
     \Im\sqbrack{\frac{E}{\Omega} \paren{1 + \frac{\Sigma}{\sqrt{\Sigma^2-4P^2}}} }, 
     \\ 
 N_{x}^{\rm } & = \Im\sqbrack{ \frac{1}{\Omega} \paren{h_{}^{} - \frac{h_{}^{}\Sigma - 2\Delta P}{\sqrt{\Sigma^2-4P^2}}} },
 \\
 N_{z}^{\rm } &= \Im\sqbrack{ \frac{1 }{\Omega} \paren{\Delta_{\rm so} - \frac{\Delta_{\rm so}\Sigma + 2\psi P}{\sqrt{\Sigma^2-4P^2}}} },  
\end{align}
\end{subequations}
where  $\Omega =\sqbrack{2\paren{\Sigma-2\rho^2 + \sqrt{\Sigma^2 - 4P^2}}}^{1/2} ,\,\, \Sigma=-E^2 + \Delta^2 +\psi^2+\Delta_{\rm so}^2+h^2,\, P = h_x^{}\Delta-\Delta_{\rm so}\psi$.

At large SOC, $\Delta_{so}\gg \Delta$, and neglecting the triplet $\psi$, the above expressions can be simplified to
\begin{subequations}
\begin{align}
   N_{0}^{\rm } &=  
     \frac{|E|}{\sqrt{E^2-\tilde{\Delta}^2}} \theta (|E|-\tilde{\Delta}), 
     \\ 
 N_{x}^{\rm } & =  \frac{h \tilde{\Delta}^2}{\rho^2}\frac{\text{sgn}(E)}{\sqrt{E^2-\tilde{\Delta}^2}} \theta (|E|-\tilde{\Delta}),
 \\
 N_{z}^{\rm } &=  -\frac{h^2 \tilde{\Delta}^2}{\Delta_{so} \rho^2}\frac{\text{sgn}(E)}{\sqrt{E^2-\tilde{\Delta}^2}} \theta (|E|-\tilde{\Delta}).  
\end{align}
\label{Eq:DOSSimp}
\end{subequations}
Placing these expressions into Eq.~\eqref{eq:TEs}, and solving the energy integral for $\Delta \gg T$ \cite{Ozaeta.2014.predicted}, we obtain Eq.~\eqref{Eq:ThermoAN}. Similarly, placing Eq.~\eqref{Eq:DOSSimp} into the expression for the charge current \eqref{eq:averaged-currents}, we calculate analytically the $I(V)$ curve at small voltages \cite{ilic2022current}, using which we then obtain rectification coefficient from Eq.~\eqref{Eq:Rect}.
%
%
%
\subsection{Material X}
We consider a generalized scenario where X can be a normal metal (NM), ferromagnet (FM), valley-polarized (VP) or spin-valley polarized (SVP, e.g. QM) material. The Hamiltonian~\eqref{eq:Hamiltonian-QM} in the main text covers all these scenarios: NM  when $m_z^{}=m_x^{} =m_{\rm v}^{} =0$, FM  when $m_{\rm v}^{} =0$, $m_z^{}\neq 0$, and/or $m_x^{} \neq 0$, VP when $m_{\rm v}^{} \neq 0$, $m_z^{}=m_x^{}= 0$, and finally for the SVP explored in the text  when $m_{\rm v}^{} \neq 0$, $m_{z}^{}\neq 0$ or $m_{x}^{}\neq 0$. 
%
%
%
\subsection{Generalized tunneling barrier}\label{App:Sec:Tunneling-Matrix}
Since the applied in-plane field and the Ising field are not aligned and X can be any material, we also consider a general scenario where the barrier could be an ordinary insulator (I), spin filter (S), valley filter (V), or a spin-valley filter (SV).  Thus, the tunneling amplitudes,  are the elements of the generalized tunneling matrix~\cite{Awoga.2026.Nonequilibrium}
\begin{equation}\label{App:eq:Tunneling-Matrix}
\begin{split}
  \hat{ \mathcal T} = &t + t^\prime\, \hat \eta_3^{} +  \bs u \cdot  \bs s +  u^\prime   \hat s_3^{}\, \hat\eta_3^{} = \hat T \otimes \hat \eta_0^{} + \hat T^\prime\otimes \hat \eta_3^{},
    \\
    \bs u = & u \paren{\cos \paren{\varphi}\sin \paren{\theta},\sin \paren{\varphi}\sin \paren{\theta},\cos \paren{\theta}}
    \end{split}
\end{equation}
where $t$ and $u$ are the usual spin-independent and spin-dependent tunneling amplitudes~\cite{Bergeret.2012.Electronic}. We introduce $ t^\prime$ and $u^\prime$ to denote, respectively, valley-dependent, and spin-valley-dependent tunneling amplitudes. The angles $\theta$ and $\varphi$ are the polar and azimuthal angle that define the polarization direction in the barrier. We define the polarization elements in the barrier as
\begin{equation}\label{eq:App_Polarization}
\begin{split}
      P_{\alpha \beta}^v\paren{\theta,\varphi} & =\frac{ \sqbrack{ \hat {\mathcal{T}} \paren{\theta,\varphi} \hat {\mathcal{T}}^\dagger \paren{\theta,\varphi}}_{\alpha \beta,v}}{\Tr \sqbrack{\hat {\mathcal{T}}^\dagger \paren{\theta,\varphi}\hat {\mathcal{T}} \paren{\theta,\varphi}}}.
    \end{split}
\end{equation}
%
%
%
\subsection{Tunneling current}
Following the standard tunneling approach~\cite{mahan.2000.many-particle}, the spin-valley resolved current is given by $I_{\sigma v} = \langle \dot N_{\sigma v}^{\rm X} \rangle$.  After some algebra, we obtain the generalized spin-valley-resolved current as
\begin{equation}\label{App:eq:Tunnling-Current}
\begin{split}
I_{\sigma v}^{\rm } \paren{\theta,\varphi} = &
 \sum_{\sigma^\prime}P_{\sigma \sigma^\prime}^{v}\paren{\theta,\varphi} \, G_{\sigma^\prime v}\times \\
 &\int_{-\infty}^\infty {\rm d}E
\, N_{\sigma}^{\rm S} \left(E \right) \sqbrack{ f_{\rm X}^{}\paren{E} - f_{\rm S}^{}\paren{E}}
\end{split}
\end{equation}
where we have defined spin-valley polarized tunneling conductance
$G_{\sigma v}=\pi e^2\,\Tr\sqbrack{\hat {\mathcal{T}}^\dagger\hat {\mathcal{T}}}\,N_{\sigma v}^{\rm X}\paren{\epsilon_{}^{}}\,N_{}^{\rm S}\paren{\epsilon}$
with
$\epsilon$ being the normal state energy, $N_{\sigma v}^{\rm X}$ the spin-valley resolved DOS of X, and $N_{}^{\rm S}$ the normal state DOS of the Ising material. Similarly, the heat currents, $\dot Q_{\sigma v}^{}$, are obtained by inserting $\paren{E-u}$ into the integrand of the corresponding current in Eq.~\eqref{App:eq:Tunnling-Current}. 

In the basis $\paren{\uparrow1,\downarrow1,\uparrow\bar1,\downarrow\bar1}$,
we collect 
$G_{\sigma v}^{ }$ and $I_{\sigma v}^{ }$
into matrices $\hat G$ and $\hat {\mathcal{I}}$, respectively. We define the total tunneling conductance,
$\mathcal G_{\rm T}^{\rm }=\Tr\sqbrack{\hat G}/4$,
and multiple polarizations
$\mathcal M_{ij}^{}=\Tr\sqbrack{\hat s_i^{}\hat \eta_j^{}\hat G}/4 \mathcal G_{\rm T}^{\rm },\, i=0,1,2,3,\, j =0,3$. 
Similarly, we define averaged currents as
$I_{ij}^{} = \tfrac{1}{4} \Tr\sqbrack{\hat s_i^{} \hat \eta_j^{}\,  \hat{\mathcal{I}} },\,  i=0,1,2,3,\, j =0,3$, from which we obtain charge (c), spin (s), valley (v), and spin-valley (sv) currents.

Analogous to the main text, in the linear regime we obtain
\begin{equation}\label{App-eq:Onsager}
   \begin{pmatrix}
    I_\lambda^{}   
        \\
        \dot{Q}_\lambda^{} 
    \end{pmatrix}   = \hat{\mathbb{R}}^{\lambda}
    \,
    \begin{pmatrix}
        V
        \\
        \delta T/T
    \end{pmatrix}, \\ 
\end{equation}
Here $\hat {\mathbb{R}}^\lambda$ with $\lambda ={\rm c,v,s,sv}$ are the response matrices. We note that other transport probes, valley, spin, and spin-valley currents can access the novel thermoelectric, $\alpha_z^{}$, and rectification, $\mathcal R_z^{}$, effects  even in basic setups. However, in the main text we focus on the charge current since it is the most easily accessible.
%
%
\subsection{Spin-valley polarized X-Insulator-Ising superconductor setup}\label{app:sec:QM-I-Ising}
In this section, we focus on the ISC/tunnel barrier/spin-valley polarized X junction presented in the main text. Also, we focus on only $0$ and $3$ current components. For spin- and valley-inactive tunneling, the barrier polarization, Eq.~\eqref{App:eq:Tunneling-Matrix}, $P_{\alpha\beta}^{}=1$.

In this case, the spin
$\paren{\mathcal M_{30}^{} \equiv \mathcal M_{\rm s}^{}}$,
valley ,
$\paren{\mathcal M_{03}^{} \equiv \mathcal M_{\rm v}^{}}$,
and spin-valley, 
$\paren{\mathcal M_{33}^{}\equiv \mathcal M_{\rm sv}^{}}$, 
polarizations are,
\begin{subequations}\label{App-eq:Polarizations}
\begin{align}
\mathcal M_{\rm s}^{} & =   \frac{ \sqbrack{\paren{G_{\uparrow 1}^{\rm }- G_{\downarrow  1}^{\rm }} + \paren{G_{\uparrow  \bar 1}^{\rm } - G_{\downarrow \bar 1}^{\rm }}}
}{4\, \mathcal G_{\rm T}^{\rm } }, 
\\
 \mathcal  M_{\rm v}^{} & =  \frac{ \sqbrack{\paren{G_{\uparrow 1}^{\rm }+G_{\downarrow  1}^{\rm }}-\paren{G_{\uparrow  \bar 1}^{\rm }+G_{\downarrow \bar 1}^{\rm }}}
 }{4 G_{\rm T}^{\rm } },
 \\
\mathcal  M_{\rm sv}^{} &  = \frac{ \sqbrack{\paren{G_{\uparrow 1}^{\rm }-G_{\downarrow  1}^{\rm }}-\paren{G_{\uparrow  \bar 1}^{\rm }-G_{\downarrow \bar 1}^{\rm }}}
}{4\, \mathcal{G}_{\rm T}^{\rm } },
\\
 \mathcal  G_{\rm T}^{} & =   \frac{ \sqbrack{\paren{G_{\uparrow 1}^{\rm }+G_{\downarrow  1}^{\rm }}+\paren{G_{\uparrow  \bar 1}^{\rm }+G_{\downarrow \bar 1}^{\rm }}}
}{4 } 
\end{align}
\end{subequations}
Assuming $\epsilon \equiv \mu \gg m_{\rm x,z,v}$, see Eq.~\eqref{eq:Hamiltonian-QM} in the main text, we expand $G_{\sigma v} $ up to quadratic term to arrive at the polarization estimations presented in the main text, providing clearer insight to the underlying mechanism.

When the polarization of X is along $z$, for $m_{\rm v,z} \neq 0$ and $m_{\rm x} = 0$, we obtain the charge
$\paren{I_{00}^{}=I_{\rm c}^{} \equiv I}$,
valley 
$(I_{03}^{}=I_{\rm v}^{})$,
spin 
$(I_{30}^{}=I_{\rm s}^{})$
and spin-valley
$(I_{33}^{}=I_{\rm{sv}}^{})$ currents as follows
\begin{subequations}\label{App-eq:averaged-currents-Mz}
\begin{align}
I_{\rm }^{}  & =  \frac{ \mathcal G_{\rm T}^{\rm }}{e}
\, \int_{-\infty}^\infty {\rm d} E\, \sqbrack{ N_{0}^{} +  \mathcal M_{\rm sv}^{z}\, N_{z}^{} }  \, f_{\rm XS}^{}\paren{E} ,
\\
I_{\rm v}^{} & =   \frac{\mathcal G_{\rm T}^{\rm }}{e}
\, \int_{-\infty}^\infty {\rm d} E\, \sqbrack{\mathcal M_{\rm v}^{z}\, N_{0}^{} + \mathcal M_{\rm s}^{z}\, N_{z}^{} }  \, f_{\rm XS}^{}\paren{E},
\\
I_{\rm s}^{} & =   \frac{\mathcal G_{\rm T}^{\rm }}{e}
\, \int_{-\infty}^\infty {\rm d} E\, \sqbrack{\mathcal M_{\rm s}^{z}\, N_{0}^{} + \mathcal M_{\rm v}^{z}\, N_{z}^{} }  \, f_{\rm XS}^{}\paren{E}, 
\\
I_{\rm sv}^{} &  =   \frac{\mathcal G_{\rm T}^{\rm }}{e}
\, \int_{-\infty}^\infty {\rm d} E\, \sqbrack{  \mathcal M_{\rm sv}^{z}\, N_{0}^{}+  N_{z}^{} }  \, f_{\rm XS}^{}\paren{E},
\end{align}
\end{subequations}
while for the polarization of X is along $x$, for $m_{\rm v,x} \neq 0$ and $m_{\rm z} = 0$, we obtain,
\begin{subequations}\label{App-eq:averaged-currents-Mx}
\begin{align}
I_{\rm }^{}  & =  \frac{\mathcal G_{\rm T}^{\rm }}{e}
\, \int_{-\infty}^\infty {\rm d} E\, \sqbrack{ N_{0}^{} + \mathcal M_{\rm s}^{x}\, N_{x}^{} }  \, f_{\rm XS}^{}\paren{E} ,
\\
I_{\rm v}^{} & =   \frac{\mathcal G_{\rm T}^{\rm }}{e}
\, \int_{-\infty}^\infty {\rm d} E\, \sqbrack{\mathcal M_{\rm v}^{x}\, N_{0}^{} + \mathcal M_{\rm sv}^{x}\, N_{x}^{} }  \, f_{\rm XS}^{}\paren{E},
\\
I_{\rm s}^{} & =   \frac{\mathcal G_{\rm T}^{\rm }}{e}
\, \int_{-\infty}^\infty {\rm d} E\,  \mathcal M_{\rm v}^{x}\, N_{z}^{}   \, f_{\rm XS}^{}\paren{E}, 
\\
I_{\rm sv}^{} &  =   \frac{G_{\rm T}^{\rm }}{e}
\, \int_{-\infty}^\infty {\rm d} E\,  N_{z}^{}   \, f_{\rm XS}^{}\paren{E}. 
\end{align}
\end{subequations}
Combining  Eq.~\eqref{App-eq:averaged-currents-Mz}(a) and Eq.~\eqref{App-eq:averaged-currents-Mx}(a) we arrive at Eq.~\eqref{eq:averaged-currents} in the main text. The superscript of the polarizations denote the orientation of spin polarization in X.  It is important to note that the spin and spin–valley currents in Eq.~\eqref{App-eq:averaged-currents-Mx}(c,d)  differ from those in  Eq.~\eqref{App-eq:averaged-currents-Mz}(c,d).  This distinction arises because, in both cases, we have specifically considered the spin-$z$ component of the current. In contrast, for the latter case, evaluating the spin-$x$ current yields spin and spin–valley currents analogous to Eq.~\eqref{App-eq:averaged-currents-Mz}(c,d)~\cite{Awoga.2026.Nonequilibrium}.

\begin{figure}[!t]
    \centering
        \includegraphics[width=0.49\textwidth]{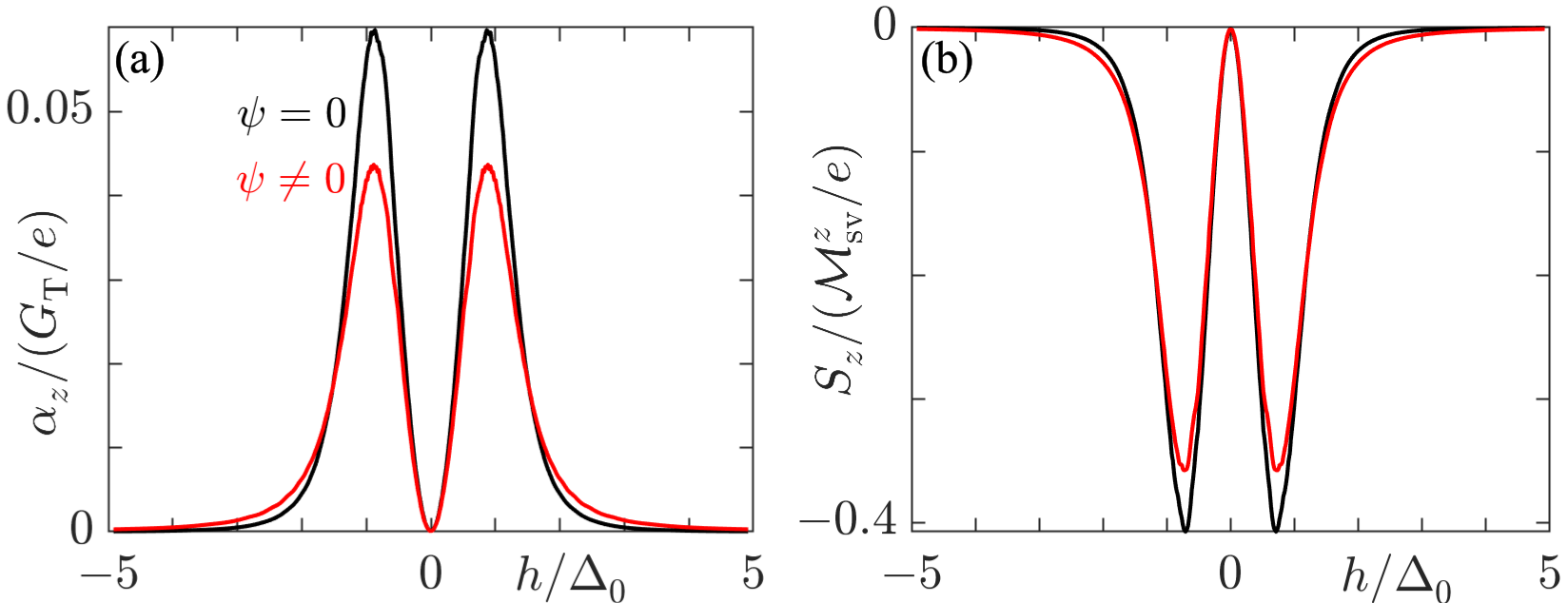}
	\caption{Effect of spin-triplet pairing, $\psi$, in Ising superconductor. (a) Ising thermoelectric, and (b) Seebeck coefficients. Here $\DeltaSO/\Delta_0=1$ and $k_{\rm B}^{}T/\Delta_0 = 0.3$, $T_{\rm ct}^{}=0.1 T_{\rm cs}^{}$.}
	\label{fig:App_TE}
\end{figure}

While the main text focuses on $\psi = 0$, here, we include finite $\psi$ for completeness. Figure.~\ref{fig:App_TE} shows that $\psi$ suppresses $\alpha_z,\, S_z$ only near the peak. This is expected, because the mirage gap contributes appreciably to the transport coefficients primarily in this region, and $\psi$ reduces the mirage gap. 
%
%
\subsection{Valley polarized X-Ferromagnetic Insulator-Ising superconductor setup}\label{app:sec:VP-SF-Ising}
We now consider a setup where a spin-filtering barrier, for instance a ferromagnetic insulator, is sandwiched between a valley polarized X and the ISC. Rhombohedrally-stacked graphene and twisted bilayer graphene are known to host valley polarized states~\cite{Awoga2024Superconductivity,chatterjee2022inter}. 
Therefore, the setup could be obtained by replacing X and the barrier with the appropriate material in Fig.~\ref{fig:Schematic}(a) in the main text.  As before, we focus on only $0$ and $3$ current components. We consider the polarization of the barrier to lie along $z$, such that $\theta =\phi =0$. From Eq.~\eqref{App:eq:Tunneling-Matrix}, we obtain the polarization components as $P_{\sigma \sigma}^{\eta}=P_{\sigma \sigma}^{}=1 + \zeta P_{\rm s}^{}$, where $\zeta =\pm 1$ for $\sigma=\uparrow/\downarrow$. We have followed the standard definition $t^2 + u^2 =1$ and $P_{\rm s}^{}=2ut$, which denotes spin polarization in the barrier~\cite{Bergeret.2012.Electronic}. 

We obtain the averaged current for this set up as,
\begin{subequations}\label{App-eq:averaged-currents-Mv}
\begin{align}
I_{\rm }^{}  & =  \frac{ \mathcal G_{\rm T}^{\rm }}{e}
\, \int_{-\infty}^\infty {\rm d} E\, \sqbrack{ N_{0}^{} +  P_{\rm s}^{}\,\mathcal M_{\rm v}^{z}\, N_{z}^{} }  \, f_{\rm XS}^{}\paren{E} ,
\\
I_{\rm v}^{} & =   \frac{\mathcal G_{\rm T}^{\rm }}{e}
\, \int_{-\infty}^\infty {\rm d} E\, \sqbrack{\mathcal M_{\rm v}^{z}\, N_{0}^{} +  P_{\rm s}^{}\, N_{z}^{} }  \, f_{\rm XS}^{}\paren{E},
\\
I_{\rm s}^{} & =   \frac{\mathcal G_{\rm T}^{\rm }}{e}
\, \int_{-\infty}^\infty {\rm d} E\, \sqbrack{ P_{\rm s}^{}\, N_{0}^{} + \mathcal M_{\rm v}^{z}\, N_{z}^{} }  \, f_{\rm XS}^{}\paren{E}, 
\\
I_{\rm sv}^{} &  =   \frac{\mathcal G_{\rm T}^{\rm }}{e}
\, \int_{-\infty}^\infty {\rm d} E\, \sqbrack{ P_{\rm s}^{}\,\mathcal M_{\rm v}^{z}\, N_{0}^{}+  N_{z}^{} }  \, f_{\rm XS}^{}\paren{E}.
\end{align}
\end{subequations}
We note that the currents, Eq.~\eqref{App-eq:averaged-currents-Mv}, are qualitatively the same as those of Eq.~\eqref{App-eq:averaged-currents-Mz} but with $\mathcal M_{\rm sv}$ replaced by $P_{\rm s}^{} \mathcal{M}_{\rm v}$ and $\mathcal{M}_{\rm s}$ by $P_{\rm s}^{}$. It is then clear that spin- and valley-polarized states can be probed in this set up. It is worth mentioning that ${\mathcal G_{\rm T}^{\rm }}$, $\mathcal M_{\rm v}$ and $\mathcal M_{\rm s}$ are  given by Eq.~\eqref{App-eq:Polarizations} but with $G_{\uparrow v}^{\rm }=G_{\downarrow  v}^{\rm }$. 
\end{document}